\documentstyle [12pt] {article}

 \newcounter{abceqn}

 \newcounter{abcfig}

\newcommand{\JFM}{J. Fluid Mech.}   
\newcommand{\JSC}{J. Sc. Comp.}   
\newcommand{\JCP}{J. Comput. Phys.}

\newcommand{\DEL}{\mbox{$\nabla$}}

\newcommand{\LAPL}{\mbox{$\nabla^{2}$}}

\newcommand{\VI}{\mbox{$\bf {v}$}}

\newcommand{\dm}{\mbox{$\dot{m}$}}

\newcommand{\bn}{\mbox{$\bf {n}$}}
\newcommand{\br}{\mbox{$\bf {r>>10}$}}
\begin {document}

%
\input psfig.sty
\psfull
\def\figurepath {Figures/}

\thispagestyle {empty}

\large
\title{Symmetry Breaking, Anomalous Scaling and Large-Scale Flow 
Generation in a Convection Cell}
\author{Ananias G. Tomboulides$^1$ and Victor Yakhot \\
Department of Aerospace and Mechanical Engineering \\
Boston  University \\
110 Cummington Street, Boston, MA 02215}

\maketitle
\centerline {A manuscript submitted to {\it Journal of Fluid Mechanics}}
\footnote{Corresponding author}

\normalsize
\begin{abstract}
We consider a convection process in a thin loop. At $Ra=Ra^{\prime}_{cr}$ a first
transition leading to the generation of corner vortices is observed.
At higher $Ra$ a coherent large-scale flow, which persists for a very 
long time, sets up. The mean velocity $\bar{v}$, mass flux $\dm$, and the Nusselt 
number $Nu$ in this flow scale with $Ra$ as $\bar{v} \propto \dot{m} \propto Ra^{0.45}$ 
and $Nu\propto Ra^{0.9}$, respectively. The time evolution of the coherent flow is well 
described by the Landau amplitude equation within a wide range of $Ra$-variation. The 
anomalous scaling of the mean velocity, found in this work, resembles the one experimentally 
observed in the ``hard turbulence'' regime of Benard convection. A possible relation between 
the two systems is discussed. 
\end{abstract}


\section{Introduction}
Thermal convection in a Benard cell is one of the classic, well-controlled, 
systems, on which one can test the theoretical understanding of various 
natural phenomena like fluid instabilities, transition,
strong turbulence itself and the laws governing heat and mass transfer 
in turbulent flows. Since convection is one of the most commonly occuring 
phenomena in Nature, the ability to describe it is also of great 
practical interest. The linear stability of an infinite fluid layer between 
two plates heated from below was investigated more than a Century ago and 
the appearence of convection rolls has since been well understood. More recent 
results on transition to turbulence in Benard cells demonstrated a beautiful 
and intricate picture of chaos onset from ordered and coherent rolls.
Studies of high $Ra$-number turbulence in a convection cell were typically  
based on the idea that the temperature profile averaged over horizontal
planes $\Theta(z)=\overline{T(x,y,z)}$ differs from an almost-constant-value 
only within thin thermal boundary layers of width $\delta_{T}$. Then, 
assuming that the upper and lower boundary layers do not ``communicate'' one 
obtains on dimensional grounds:

\begin{equation}
\delta_{T}\approx Ra^{-\frac{1}{3}}
\end{equation}

\noindent 
leading to the scaling of the dimensionless heat flux (Nusselt number, defined below)

\begin{equation}
Nu\approx Ra^{\frac{1}{3}}
\end{equation}

The accurate experiments on Benard convection, using low temperature 
helium, conducted by the Libchaber group, Heslot {\it et al.} (1987), 
Castaing {\it et al.} (1989), showed that this relation is, in fact, 
incorrect and instead the Nusselt number at $Ra\geq 10^{8}$ scales as:

\begin{equation}
Nu\approx Ra^{x}
\end{equation}

\noindent 
with $x\approx 0.285$, which is close to $x=2/7$. The experiments
demostrated the appearence of this scaling as a transition from 
a random state of the fluid with the heat transfer dominated by 
small-scale vortical motions, observed at $10^{3}<Ra<10^{8}$, to a 
new state of turbulence, characterized by the onset of a powerful and
persistent coherent large scale-flow (boundary layer ``wind''). 
This wind, fascilitating strong correlation of the top and bottom 
boundary layers, leads to deviations from the ``classic'' scaling 
exponent $\xi=1/3$. 

Since the first experiments, Heslot {\it et al.} (1987), Castaing 
{\it et al.} (1989), this effect was observed in Benard convection with 
air, helium, water and mercury as working fluids, different aspect ratio 
cells etc, Wu (1991), Wu and Libchaber (1992), Belmonte {\it et al.} (1994).
The experiments revealed that the dependence of the ``wind'' velocity with 
$Ra$ is:

\begin{equation}
V\propto Ra^{\xi}
\label{eqn.V}
\end{equation}

\noindent 
with $\xi\approx 0.48-0.49$ differing from the expected free-fall
exponent $\xi=0.5$. 

The proposed theoretical models mainly dealt with the explanation of 
the observed $x=2/7$ exponent, assuming the existence of the large scale 
flow. Understanding the reasons for the appearence of 
coherent flow in strongly turbulent Benard convection was 
the prime motivation of this work. However, the unexpected results,
presented below, are of an independent interest disregarding their 
relation to turbulent convection.

Our motivation was prompted by the following qualitative argument.
We consider a fluid with Prandtl number, $Pr=\nu_0/\kappa_0 \approx 1$,
where $\nu_{o}$ and $\kappa_{o}$ stand for viscosity and thermal diffusivity.
Assume that the flow in the pre-transition (``soft turbulence'') state is simply  
a traditional Kolmogorov-like turbulence. We treat the flow as consisting of 
two parts: a viscous boundary layer of width $\delta\approx \delta_{T}$ and the 
bulk, where the effective transport coefficients are estimated as:
 
\begin{equation}
\kappa\approx \nu \approx u_{rms}L = \frac{u_{rms}L}{\nu_0} {\nu_0} \approx \nu_{o}Re
\end{equation}

Thus, the bulk can be perceived as a very viscous (large mass) fluid.
Setting, for the sake of the argument, $\nu\rightarrow \infty$ we 
conclude that the problem of stability of the thin boundary layer adjacent 
to the walls of the convection cell can be decoupled from the ``super-stable'' 
chaotic flow in the bulk. A somewhat similar situation was considered in the 
end of the sixties by Welander (1967) and Keller (1967).
interested in a convection process in a fluid contained within a thin loop 
heated from below. It has been shown that when $Ra$ was large enough 
this system is unstable and a steady clock-wise or counter-clockwise flow 
sets up. In some situations an oscillatory solution was possible.

The physical explanation of this effect is as follows: The mean bouyancy force in 
this system is approximately, Welander (1967):

\begin{equation}
\bar{B} \approx A g\rho_{o}\alpha \int T dy
\label{eqn.B}
\end{equation}

\noindent where $A$ is a cross-sectional area, $g$ is the gravity
acceleration, $\alpha$ stands for the thermal expansion coefficient, $T$ for a 
temperature distribution and $y$ is a coordinate in the vertical direction. 
The bouyancy is balanced by a mean friction force $\bar{F}$, depending on the flow
rate $q=\bar{v} A$ (where $\bar{v}$ is the mean velocity)

\begin{equation}
\bar{F}\propto q\propto \bar{v}
\label{eqn.F}
\end{equation}

\noindent when $q$ is not too large. The most important part of the argument
is that the buoyancy $\bar{B}=\bar{B}(q)$. Model equations, developed in Welander 
(1967) and Keller (1967),
showed that $\bar{B}(q)$ was only weakly $q$-dependent and,  as a consequence, the
curves $\bar{B}(q)$ and $\bar{F}(q)$ must cross at least at one point. That is the reason
why a regular flow sets up in this system.

Thus, the qualitative picture of turbulence in Benard cell, 
presented above, combined with an idea that the low viscosity boundary layer
is, in some respect, decoupled from the bulk makes the analogy with
convection in a thin loop possible. However, if this analogy holds, it must
explain the experimentally observed anomalous scaling $V(Ra)$, from ~(\ref{eqn.V}).  
This is a prime goal of this paper.

We investigated the dynamics of the flow in two-dimensional cells $0 \leq x,y\leq L$ 
with viscosity $\nu=\nu_{0}\leq \infty$ in the interval $\delta \leq |x|,|y| \leq L-\delta$.
Outside this interval $\nu=\infty$ meaning that $\VI=0$. Experiments showed that 
the main contribution to the heat transfer (more than $50\%$) came from the
wind. That is why at this stage we neglected the heat tarnsfer in the 
inner part of the cell, setting there $\kappa=0$.

In order to test the sensitivity of the results to the geometry of the
system, we have performed simulations in three configurations, corresponding to cells 
of different geometry. The first configuration is a cell of $L=1$ and 
$\delta/L=0.1$, the second is identical with the first, except for the aspect
ratio $\delta/L=0.05$ instead of $0.1$, whereas the third one corresponds to
a circular geometry with the same diameter $L$ and aspect ratio with the first
case; in the latter case, the bottom fourth of the circumference of the circle
is maintained at a high non-dimensional temperature of $1$, whereas the top fourth 
is kept at a low temperature of $0$. The three convection cells and the boundary conditions used
are shown in Fig. ~\ref{fig.config_all}I,II,III.

\section{Formulation}

The equations of motion are the Boussinesq equations, outlined below:

\begin{eqnarray}
\frac{\partial \VI}{\partial t} + \VI \cdot \DEL \VI & = &
-\DEL p + T + \frac{1}{Re} \LAPL \VI \nonumber \\
\\
\frac{\partial T}{\partial t} + \VI \cdot \DEL T & = & \frac{1}{Re Pr} \LAPL T \nonumber \\
\\
\DEL \cdot \VI & = & 0 \nonumber
\label{eqn.motion}
\end{eqnarray}

\noindent
where the non-dimensionalizing velocity scale is $U=\nu/L Gr^{1/2}$ and $Re=Gr^{1/2}$,
where $Gr$ is the Grasshof number (here, $Ra=Gr$ since $Pr=1$).
The only non-dimensional parameter in the system (except for shape and aspect ratios)
is the Rayleigh number, $Ra={\beta g \Delta T L^3}/{\kappa \nu}$.
The results from simulations we have performed will be reported as function of $Ra$,
and in particular as function of $r=(Ra-Ra_{cr})/Ra_{cr}$, where $Ra_{cr}$ is a critical 
Rayleigh number where transition to large scale mean flow occurs.

For the time integration of equations ~(\ref{eqn.motion}), we use a fractional 
step method, in conjunction with a mixed explicit/implicit stiffly stable 
scheme of second order of accuracy in time, Karniadakis {\it et al.} (1991). 
A consistent Neumann boundary condition is used for the pressure, 
based on the rotational form of the viscous term, which nearly eliminates 
splitting errors at solid (Dirichlet) velocity boundaries, Tomboulides {\it et al.} (1989).
The spatial discretization of the resulting Helmholtz equations is
performed using two-dimensional Legendre spectral elements, Patera (1984).
The resulting matrices for the numerical solution of the two-dimensional Helmholtz equations 
are solved using preconditioned conjugate gradient iterative solvers. 
The resolution can be increased by either increasing the number of elements 
or the order of the interpolants inside each element. In the simulations 
presented here, the resolution was improved mainly by increasing the order of interpolants.
Several resolution tests, not reported here, were performed and a typical discretization 
consists of $40$ elements with up to $15$ points in each direction per element.
In general, because of the laminar nature of the flow in the range of $Ra$ numbers
investigated, the computational cost was not a limiting factor; a typical simulation
took few hours on a SGI/R10000 workstation. The resolution became limiting only
in the simulation of very high $Ra$ cases (over $10^9$ or so) described in the
last section. 

\section{First transition}

At very low Rayleigh numbers, the fluid inside the cell is not in motion. As $Ra$
increases over $Ra^{\prime}_{cr}$, the first transition, corresponding to the 
appearance of convection rolls, occurs. Here, because of the geometry, 
the convection rolls at the scale $l\approx \delta$ appear only at the corners of 
the domain, and display a double-flip symmetry with respect to the diagonal. Isocontours 
of vorticity, consisting of counter-rotating vortices located at the four corners,
and for $Ra \le Ra_{cr}$ (corresponding to the second transition), are shown 
in Figures ~\ref{fig.First_transition_Vort}I,II,III.

This flow configuration is typical for all $Ra^{\prime}_{cr} < Ra < Ra_{cr}$.
This range of $Ra$ numbers was not investigated further, since 
our interests lie in the study of the large scale mean flow which appears for $Ra 
> Ra_{cr}$ when the symmetry of the flow disappears.

\section{Second transition}
As the $Ra$ number increases, a symmetry-breaking, resulting in the appearance of a large scale 
mean flow, occurs. This transition is a linear one and corresponds to a regular bifurcation 
(or exchange of stability) where the resulting flow is not time-dependent (i.e. the crossing 
eigenvalue has zero frequency). A typical behavior of the total kinetic energy of the flow 
is shown in Fig. ~\ref{fig.A_vs_t_Ra30000_transition} which corresponds to case I for 
$Ra=30,000$ ($Ra_{cr}$ for this case is equal to $28,289$).

Isocontours of the $x,y$ velocity components $u, v$, temperature, $T$, and vorticity $\omega$, 
are plotted for cases I, II, and III, in Fig. ~\ref{fig.Second_transition_all}. The generation
of a large scale mean flow is evident when comparing, e.g. the vorticity with Fig. ~\ref{fig.First_transition_Vort}.
The direction of the flow (clock- or counter-clockwise) is random and depends on the initial 
round-off error disturbances. For example, it can be observed from the same Figure, that for 
case I the large scale flow has a clockwise rotation, whereas the opposite is true for the 
other two cases II, and III. 

The average mass flux ($\dm$) and non-dimensionalized heat flux ($Nu$) for case I are plotted in 
Figures ~\ref{fig.plot_m_vs_Ra_I} and ~\ref{fig.plot_Nu_vs_Ra_I}, respectively, as functions 
of $r=(Ra-Ra_{cr})/Ra_{cr}$. The value of $Ra_{cr}$ for case I, was found to be 
$Ra_{cr}=28289$. As can be observed, the mass flux scales approximately as 
$\dm \propto r^{0.45}$ for a large range of $Ra$ numbers ($r$ between $0.1$ and $10$). However, 
very close to $Ra_{cr}$, $\dm \propto r^{0.5}$ as expected, Landau (1987).
The scaling exponent of $Nu$ also approaches $1$ as $Ra \rightarrow Ra_{cr}$.
In section ~\ref{sec.Landau}, an analysis is presented in the context of the Landau amplitude 
equation which explains the two types of behavior. It has to be noted that to obtain the steady
state value of the mass and heat flux for cases close to $Ra_{cr}$ the equations of motion
had to be integrated for very long non-dimensional times and steady state results for $r$ below
$0.01$ ($log(r)=-2$) were not performed due to the computational cost involved. This is the reason 
that one can only observe the anomalous $0.45$ scaling for case II, since the transition in the 
scaling exponent from $0.5$ to $0.45$, in this case, occurs below $r=0.01$. The variation of $\dm$ 
and $Nu$ for case II is shown in Figures ~\ref{fig.plot_m_vs_Ra_II} and ~\ref{fig.plot_Nu_vs_Ra_II}, 
respectively; here, $Ra_{cr}=114648$ and it can be seen that the $r^{0.45}$ scaling is present for 
the whole range $r$ investigated. The same holds for $Nu$ which behaves as $r^{0.9}$ 
for $0.01\le r \le 1$.

At $Ra \le Ra_{cr}$, where $\bar{v}=0$, the temperature profile is symmetric with respect to the 
plane $x=0$ and the temperature distribution is a solution to the Laplace equation; this solution
is very close to a linear profile in $y$. As flow with an average velocity $\bar{v}=\dm/\delta$ 
starts to develop, the temperature profile becomes more and more asymmetric and the temperature distribution
in $y-$ deviates from the almost linear profile at $Ra_{cr}$. As expected, as convection continues
to increase the temperature profile steepens close to $y=1$ (for clockwise rotation and along the
left vertical channel). This steepening is governed by the following equation:

\begin{equation}
v\frac{\partial T_e}{\partial y} = Ra^{-1/2} \frac{\partial^2 T_e}{\partial y^2}
\label{eqn.expT}
\end{equation}

\noindent
with boundary conditions $T=0$ at $y=0$ and $T=1$ at $y=1$.
The solution to this equation, $T_e=\left( exp(vRa^{1/2}y)-exp(vRa^{1/2}) \right) / 
\left(1-exp(vRa^{1/2}) \right)$, for different values of $vRa^{1/2}$, is compared with temperatures 
profiles along $y$ (along $x=0.05$ which corresponds to the middle of the left vertical channel) obtained 
from the numerical simulations. This comparison is shown in Fig. ~\ref{fig.profT_x0.05_vsRa} and 
as can be observed the simple model ~(\ref{eqn.expT}) can describe the steepening of the
temperature layer close to $y=1$ from $Ra=Ra_{cr}$ up to $r=10$. 

After this steepening of the temperature profile, the relation $\dm \propto r^{0.45}$ is not 
valid any more. This occurs for values of $vRa^{1/2} \ge 50$. After this, the mass flux, $\dm$, 
stops increasing and its value stays at approximately the same level. On the other hand, a mixing 
process sets up, leading to a homogenization of the temperature along the left and right 
vertical channels. The average temperature in the two channels becomes equal at very high 
$Ra \approx 10^8$, and the long time solution of the equations is not steady any 
more. It is shown in section ~\ref{sec.High_Ra} that in this range of $Ra$, the thermal energy 
input can no longer be converted to a coherent large scale flow but is almost entirely converted 
into small scale features and eventually turbulence. 

In order to explore this anomalous scaling $\dm \propto \bar{v} \propto r^{0.45}$, and to verify 
that it is not due to numerical errors, or due to numerical singularities (at corners), we analyzed 
case III which consists of a perfectly symmetric and non-singular geometry. Here, the value $Ra_{cr}$ 
was found to be equal to $34,417$. Figures ~\ref{fig.plot_m_vs_Ra_III} and ~\ref{fig.plot_Nu_vs_Ra_III} 
show the variation of $\dm$, and $Nu$ with $r$, respectively. As can be observed from these figures, 
the scaling exponents in this case are the same as the ones of case I, for both $\dm$ and $Nu$. 
For all cases it was found that $\dm \propto r^{1/2}$ for $r \leq 0.1$ or so, and 
$\dm \propto r^{0.45}$ up to $r \leq 10$. It was also found that the non-dimensionalized heat flux, 
$Nu \propto r$ for approximately $r \leq 0.1$, and $Nu \propto r^{0.9}$ for $0.1 \leq r \leq 10$. 

The results for case III are very close, both qualitatively and quantitatively, to those for 
case I. On the other hand, case II differs significantly from both cases I and III; the $\dm 
\propto r^{0.5}$ and $Nu \propto r$ range has not been observed. We believe that in this case 
this range is too narrow to be detected numerically. Thus, all we observed was the $\dm \propto 
r^{0.45}$ and $Nu \propto r^{0.9}$ range for the whole range of $r$ investigated.

\section{Scaling of Nu with Ra}

One can use an integral form of the energy equation to obtain an estimate of the
scaling of $Nu$ number with $r$. Using the steady state results for all $Ra>Ra_{cr}$,
we used a control volume which consists of the top half of the convection cell, as
shown in Figure ~\ref{fig.CV}. The energy equation integrated over this control volume is

\begin{equation}
\oint_{CV} (\bn \cdot \VI) T ds = \frac{1}{RePr} \oint_{CV} \frac{\partial T}{\partial n} ds
\label{eqn.CV}
\end{equation}

The convective heat flux is non-zero only at the lower part of the control volume along 
sides $1$ and $2$ as shown in Fig. ~\ref{fig.CV}, whereas the diffusive heat flux is only 
non-zero at the top boundary $T$ and sides $1$ and $2$. 
Also, since the vertical walls at interfaces $1$ and $2$ are insulated, the temperature
profiles in the $x-$direction are very close to constant and therefore the $x-$dependence
of $T$ at $1$ and $2$ can be neglected. One can also add and subtract the mean velocity
$\bar{u}$ (which is the same at both $1$ and $2$ due to incompressibility) and obtain
the following expression from equation ~\ref{eqn.CV} (by assuming an arbitrary clockwise
circulation):

\begin{equation}
-\dot{q}_T = \bar{u} \delta \left( \bar{T}_{1}- \bar{T}_{2} \right) - \frac{1}{RePr}
\left( \frac{\partial \bar{T}_{1}}{\partial y} + \frac{\partial \bar{T}_{2}}{\partial y} \right)
\label{eqn.CV3}
\end{equation}

\noindent
which after non-dimensionalization with the conductive heat flux $\dot{q}_{cond}$ becomes:

\begin{equation}
Nu-1 = \frac{\bar{u} \delta \left( \bar{T}_{1}- \bar{T}_{2} \right)}{\dot{q}_{cond} } +
\left[ \left( \frac{\partial \bar{T}_{1}}{\partial y} + \frac{\partial \bar{T}_{2}}{\partial y} \right) /
\left( 2\frac{\partial \bar{T}_{1}^{*}}{\partial y} \right) - 1 \right]
\label{eqn.CV4}
\end{equation}

The second term, T2, in the right hand side of equation ~(\ref{eqn.CV4}) (inside the square brackets) 
is equal to $0$ when $r=0$ (conduction regime). When $Nu >> 1$, this ${\cal{O}}(1)$ term becomes 
negligible.
Therefore, when $Nu >> 1$, one would expect a scaling of $Nu-1$ very close to the scaling of the 
first term, T1, in the right hand side.
Our results indicate that $\bar{v} \propto r^{0.5}$ for $r<0.1$, and $r^{0.45}$ for $0.1<r<10$. 
We have also found that the difference $(\bar{T}_{1}-\bar{T}_{2}) \propto r^{0.5}$ up to $r=0.1$ and 
then starts leveling off up to $r=10$ at which point it starts decreasing, due to the mixing process.
Figure ~\ref{fig.lv_ltd_Nu} shows the variation of the logarithm of these two terms, as well as the 
variation of their sum which should be equal to $Nu-1$ in a steady flow. It can be observed that both 
terms start increasing as $r^{1.0}$ at $r << 1$ and for $r \ge 1$ the second term becomes negligible. 
It can also be seen that as long as the flow at long times is steady, the sum of these two terms equals
$Nu-1$ (shown with big open circles) and its scaling is also $r^{1.0}$ for $r$ up to $0.1$ and
$r^{0.9}$ after that and up to $r=10$.

%
%
%

\section{Amplitude Equation}
\label{sec.Landau}

After the second transition at $Ra=Ra_{cr}$, a large scale mean flow is generated, the 
amplitude of which increases with $Ra$. The kinetic energy of this flow, $A^2$, integrated 
over the whole domain was used as an indicator of transition and the Landau amplitude 
equation was used to model this transition. The kinetic energy $A^2$ is governed by the
following amplitude equation:

\begin{equation}
\frac{dA^2}{dt} = \gamma A^2 - \alpha A^4
\label{eqn.Landau}
\end{equation}

The solution to this equation is given by the following expression

\begin{equation}
A^2(t) = {\gamma}\left( { \frac{\gamma-\alpha A_0^2}{A_0^2} e^{-\gamma(t-t_0)} + \alpha} \right)^{-1}
\label{eqn.Landau_soln}
\end{equation}

\noindent
where $\gamma$ and $\alpha$ are the so-called Landau constants, and $A_0=A(t=t_0)$. We found 
that our results can accurately be modelled using equation ~(\ref{eqn.Landau_soln}) for a wide
range of $Ra$ numbers much higher than $Ra_{cr}$. It is clear from ~(\ref{eqn.Landau}) that at
steady state $A$ is given by


$$A=\sqrt{\frac{\gamma}{\alpha}}$$

\noindent
and since we know the amplitude $A$ from the DNS, we varied the value of
$\gamma$ to obtain the closest description of our data. 
Figure ~\ref{fig.plot_his_vs_Landau_Ra40000_III} shows the time history of the total
kinetic energy of the flow, $A^2$, for case III and $Ra=40,000$ or $r=0.16$. The solid
line corresponds to numerical simulation results, whereas the dotted line is obtained
using equation ~\ref{eqn.Landau_soln}; as can be observed from the figure the difference
is almost negligible. This is true for $r$ up to about $10$ or so, a result which is
quite surprising when taking into acount that equation ~(\ref{eqn.Landau_soln}) is only
valid close to transition and more surprisingly that the same equation models data 
demonstrating the anomalous scaling of $r^{0.45}$ within this range of $r$.

The Landau constants $\gamma$ and $\alpha$ were calculated from numerical simulation 
results and are shown in Figures ~\ref{fig.plot_Landau_exponents_I}, 
cases I, II, and III, respectively. These figures suggest that for cases I and III considered,
$\gamma \propto r^1$ for $r \rightarrow 0$ (as expected for the kinetic energy), whereas 
$\alpha$ is approximately constant in this range and equal to approximately $100$; at about
$r=0.1$ or so the scaling exponent for $\gamma$ changes to about $0.9$ (shown with solid line)
and this scaling persists up to $r$ between $5$ and $10$. This is consistent with the 
results presented in the previous section $\dm \propto r^{0.5}$ for $r<0.1$ and $\dm \propto 
r^{0.45}$ after that. The second constant $\alpha$ stays almost constant during this transition. 

For values of $r$ higher than about $5$ or so a transition is observed in the scaling
of $\gamma \propto r^{0.9}$ scaling is no longer valid. This transition also affects 
the magnitude of the second constant $\alpha$, (which is supposed to be independent 
of $Ra$) in the interval of $Ra$-variation after the second transition, and now this
second constant starts to decrease. After this value of $r$ equation ~(\ref{eqn.Landau_soln}) 
no longer describes the data. In fact, for values of $r$ greater than $10$ one can clearly 
observe the appearance of other modes which are oscillatory and although damped at long 
times, they do appear in the initial transients. These modes lead to instabilities at $Ra$ 
numbers higher than $10^8$ for case I, and after that the flow becomes time dependent.

For case II again the $\gamma \propto r^1$ dependence was not observed. Instead, the
value of $\gamma$ was found to scale with $\gamma \propto r^{0.9}$ for $0.01<r<1$.
At $r \ge 1$ this relation breaks down, in contrast to the $\dm \propto r^{0.45}$ law 
which is valid up to $r=10$ or so. In addition, equation ~(\ref{eqn.Landau_soln}) is still 
a good approximation to the time variation of the kinetic energy for at least up to 
$r=10$.  The second exponent $\alpha$ was found to be very close to $190$ for all $r$ 
up to $1$ and to decrease monotonically for higher $r$.

The most unusual feature of the systems considered above is that the Landau
equation description, originally proposed for the immediate vicinity of the transition point, 
is remarkably accurate in the entire range $0<r<10$.

\section{Higher Ra corresponding to $\br$}
\label{sec.High_Ra}

Values of $r$ significantly higher than $10$ were investigated only for case I. It was
found that even for $Ra$ numbers at which the long time solution is steady, oscillatory
modes appear in the initial transients. In some cases these modes are similar to the
pulsating instabilities analyzed in Welander (1967), Keller (1967), where the flow may actually
reverse its direction (clock- or counter-clockwise) before it settles in a steady state. 
An example is shown in Fig. ~\ref{fig.Oscill} where the time history of the $u-$velocity
component is shown at two different points, one in the middle of the lower channel and
one in the middle of the upper horizontal channel for case I and for $Ra=10^7$.
As can be observed from this figure, the flow reverses once at a non-dimensional time 
of around $t=20$ and another time at $t=50$ before it approaches a steady state after a 
damped oscillatory behavior.

As $Ra$ increases to $Ra \ge 10^8$, the flow becomes unsteady. This unsteadiness is 
likely the result of finite amplitude disturbances. The flow turns abruptly generating the 
disturbances at the corners. However, now the viscosity is not sufficient to damp these 
disturbances and they work as a continuous source of noise which drives the channel flow unsteady, 
Patera and Orszag (1983). Results at these high $Ra$ numbers are shown in terms of vorticity 
and temperature isocontours in Figure ~\ref{fig.isocontours_I} for $Ra=10^8$ and $10^9$ 
respectively. As we can see from these Figures, the entire thermal energy input is now
transformed into finite amplitude vortical structures instead of organized large scale flow. 
The existence of these structures may give rise to three-dimensionality and eventually turbulence 
via secondary type instabilities, Patera and Orszag (1983).

\section {Summary and Conclusions}
We considered a simple system which can mimic the experimentally observed 
large-scale flow generation 
in Benard convection. This is a two-step process. First, the instability
leading to a symmetric steady flow pattern happens at $Ra=Ra_{cr}$. 
The instability of this pattern leads to a symmetry-breaking large-scale flow. Close to $Ra_{cr}$
the flow rate $ \dm \propto r^{0.5}$ and $Nu\propto r$ which is the expected
result.

After the symmetry-breaking the observed $\bar{v} \propto \dm \propto Ra^{0.45}$ and 
$Nu \propto Ra^{0.9}$. The onset of this anomalous scaling is correlated with the 
simultanious modification of the Rayleigh number dependence of the coefficients in 
the Landau amplitude equation, accurately describing the data in an extremely wide 
range of the $Ra$-number variation. The observed effect and numerical values of the
exponents seems to be insensitive to the geometric features of the system 
at least for the three cases considered in this paper. At the present time we
do not fully understand the physical origins of the anomaly. 

The qualitative similarity of the systems considered in this paper with the
high $Ra$ number Benard convection may explain the experimentally observed
large-scale flow generation leading to transition from ``soft'' to ``hard'' 
turbulence in a convection cell, Heslot {\it et al.} (1987), Castaing {\it 
et al.} (1989). This analogy is supported by the similar anomalies in the  
scaling of the ``wind'' velocity $V(Ra)$ observed in both systems. 

\section {Acknowledgements}
We are grateful to W. Malkus and E. Spiegel for bringing references Welander (1967) 
and Keller (1967) to our attention.

\newpage
\section{Figure Captions}

\begin{figure}
\centerline{\psfig{file=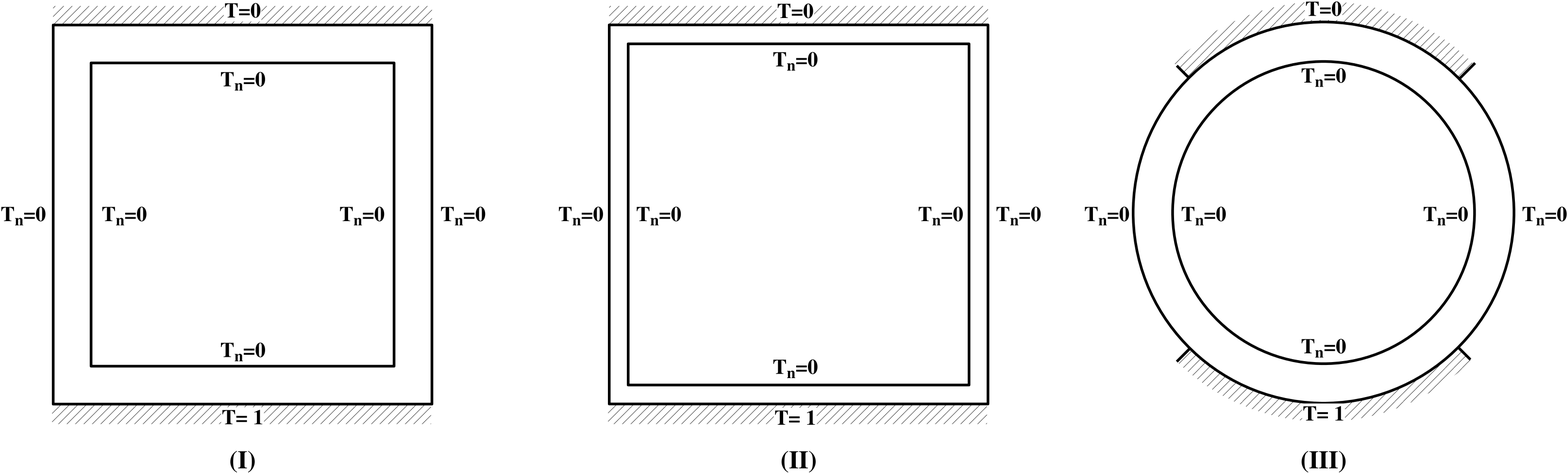,width=6.0in}}
\caption{}
\label{fig.config_all}
\end{figure}

\begin{figure}
\centerline{\psfig{file=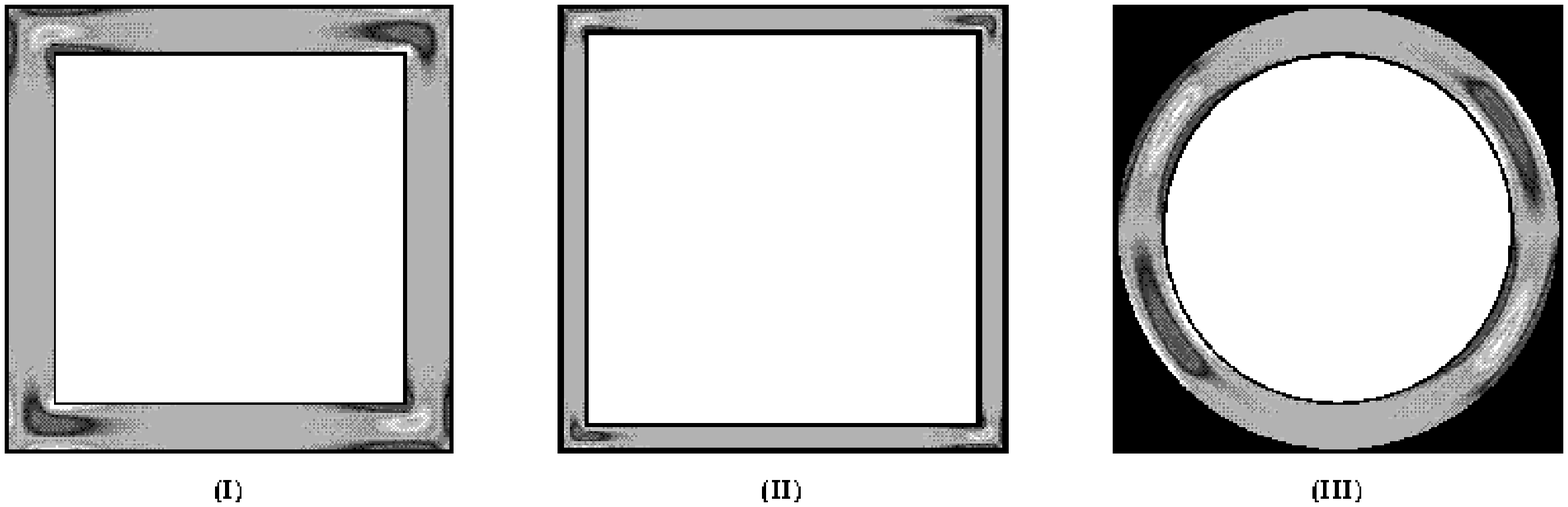,width=6.0in}}
\caption{}
\label{fig.First_transition_Vort}
\end{figure}

\begin{figure}[t]
\centerline{\psfig{file=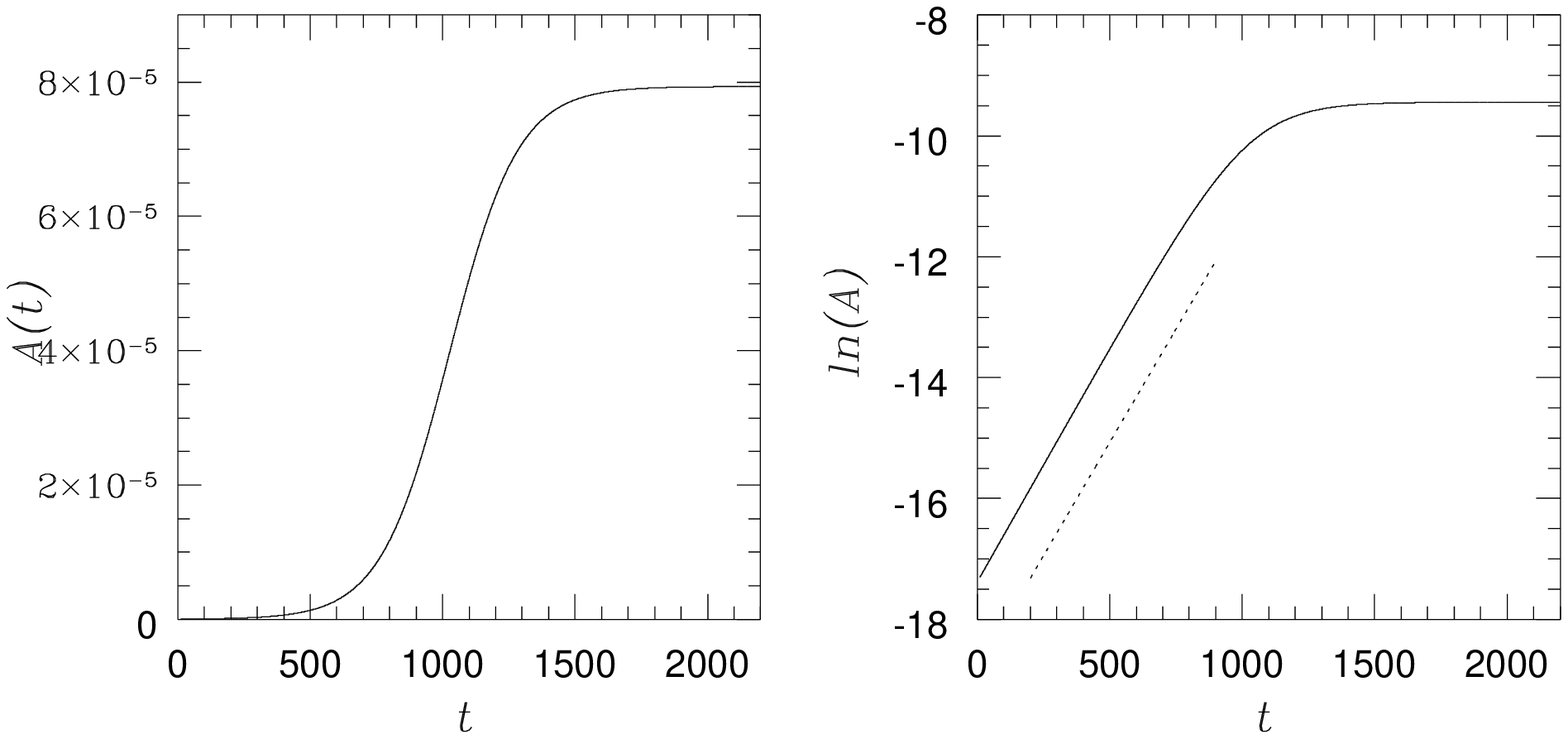,width=6.0in}}
\vspace*{-3.0in}
\caption{}
\label{fig.A_vs_t_Ra30000_transition}
\end{figure}

\begin{figure}
\centerline{\psfig{file=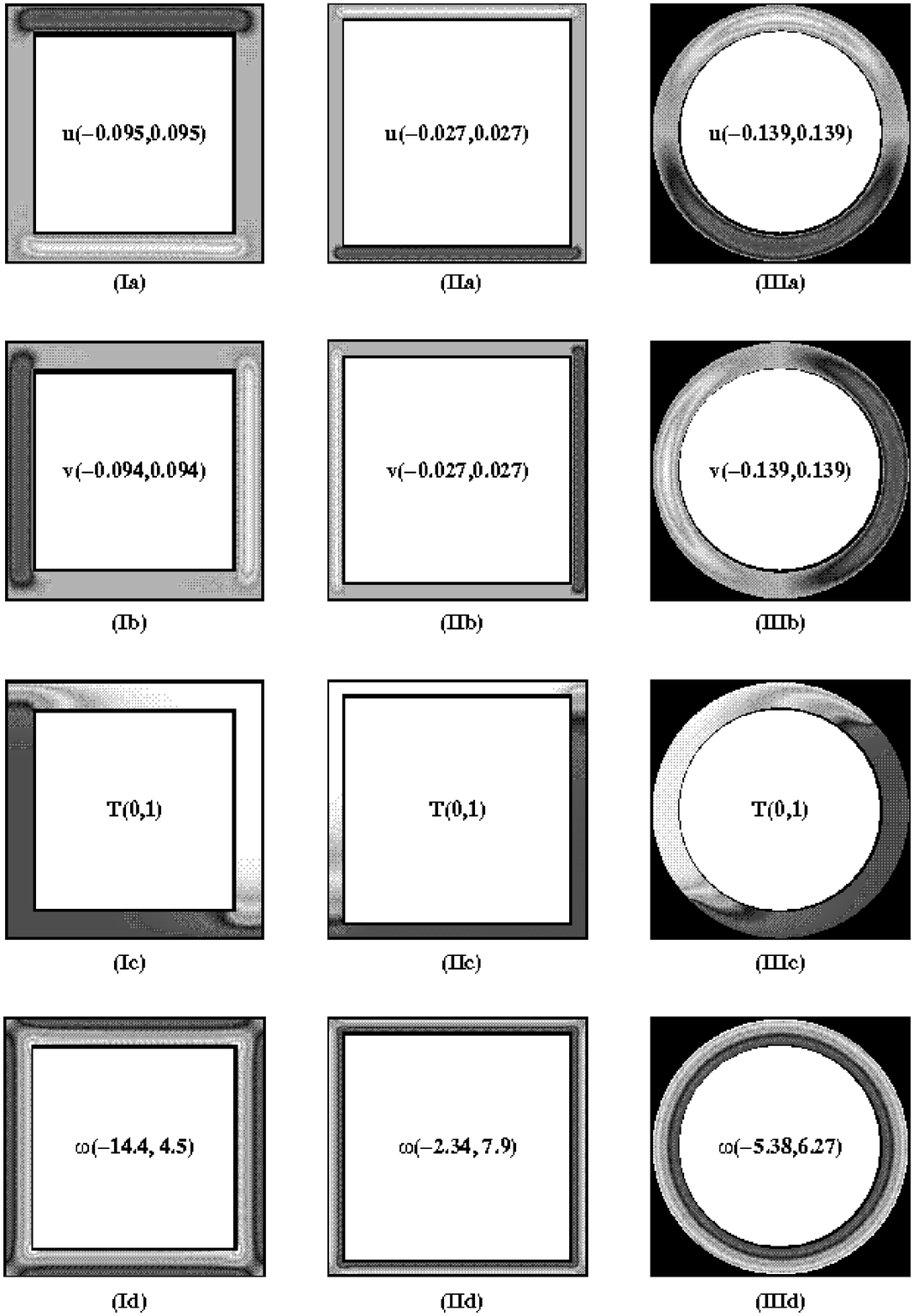,width=6.0in}}
\caption{}
\label{fig.Second_transition_all}
\end{figure}

\begin{figure}
\centerline{\psfig{file=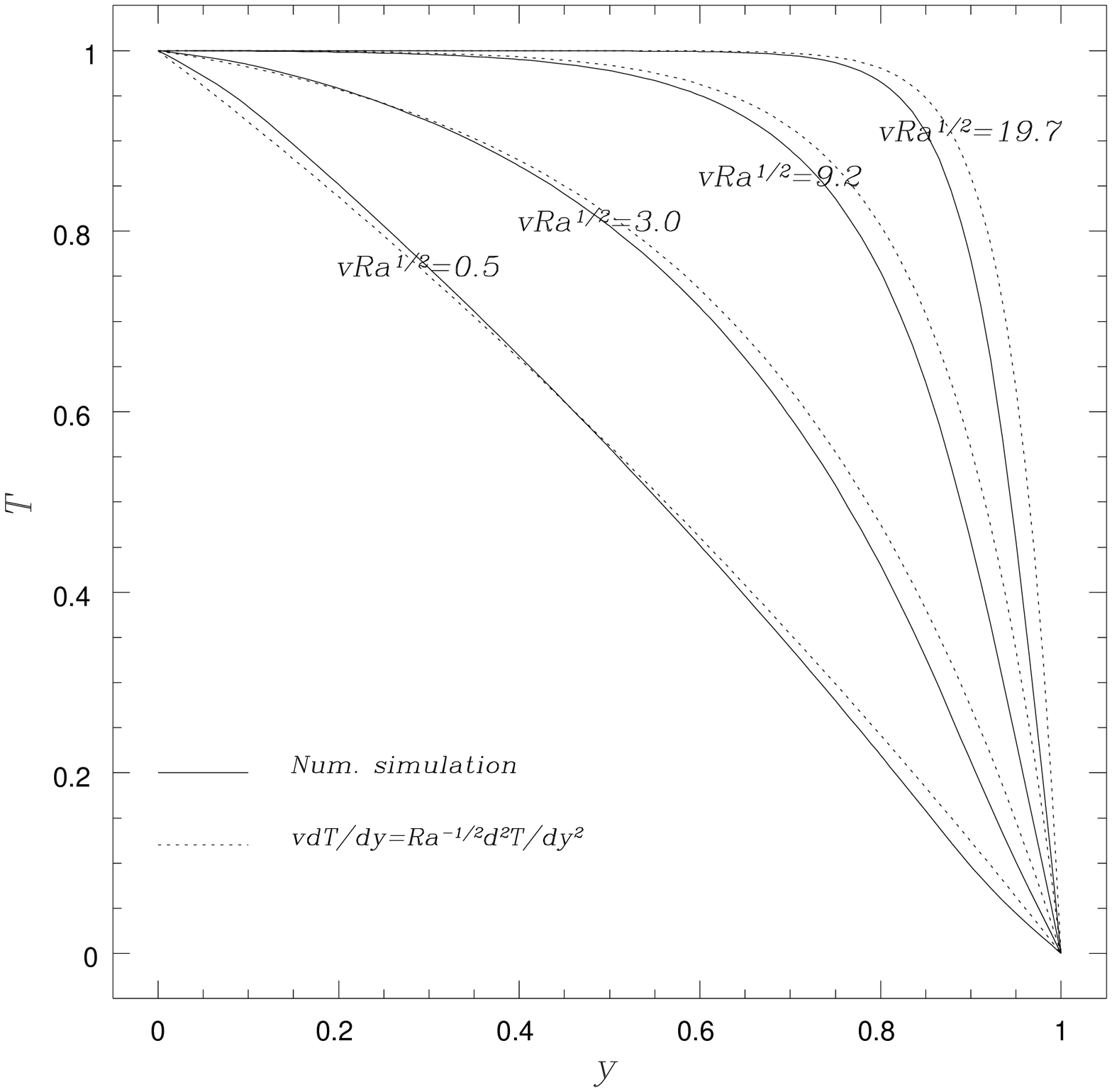,width=6.00in}}
\caption{}
\label{fig.profT_x0.05_vsRa}
\end{figure}

\begin{figure}
\centerline{\psfig{file=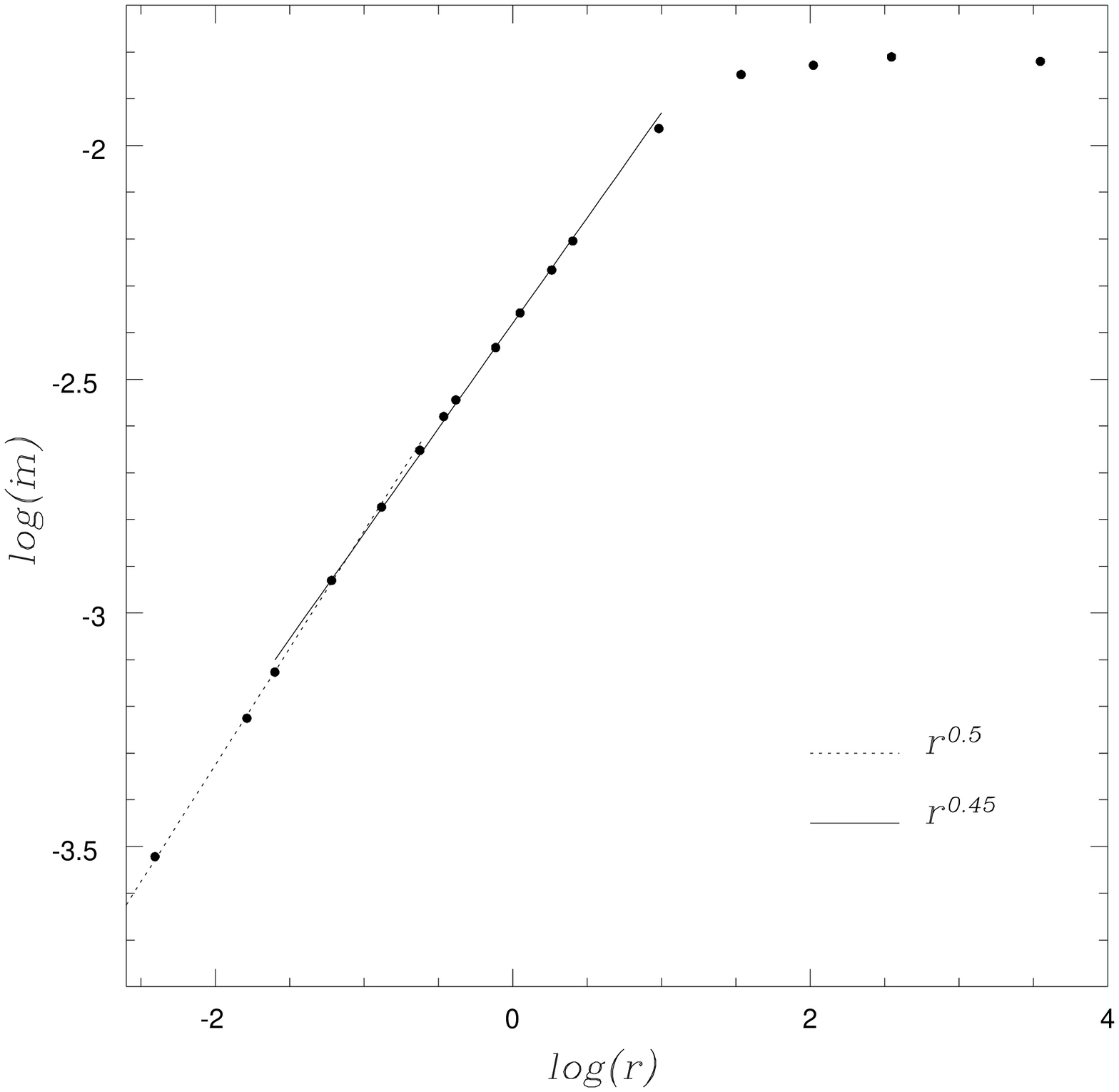,width=3.75in}}
\caption{}
\label{fig.plot_m_vs_Ra_I}
\end{figure}

\begin{figure}
\centerline{\psfig{file=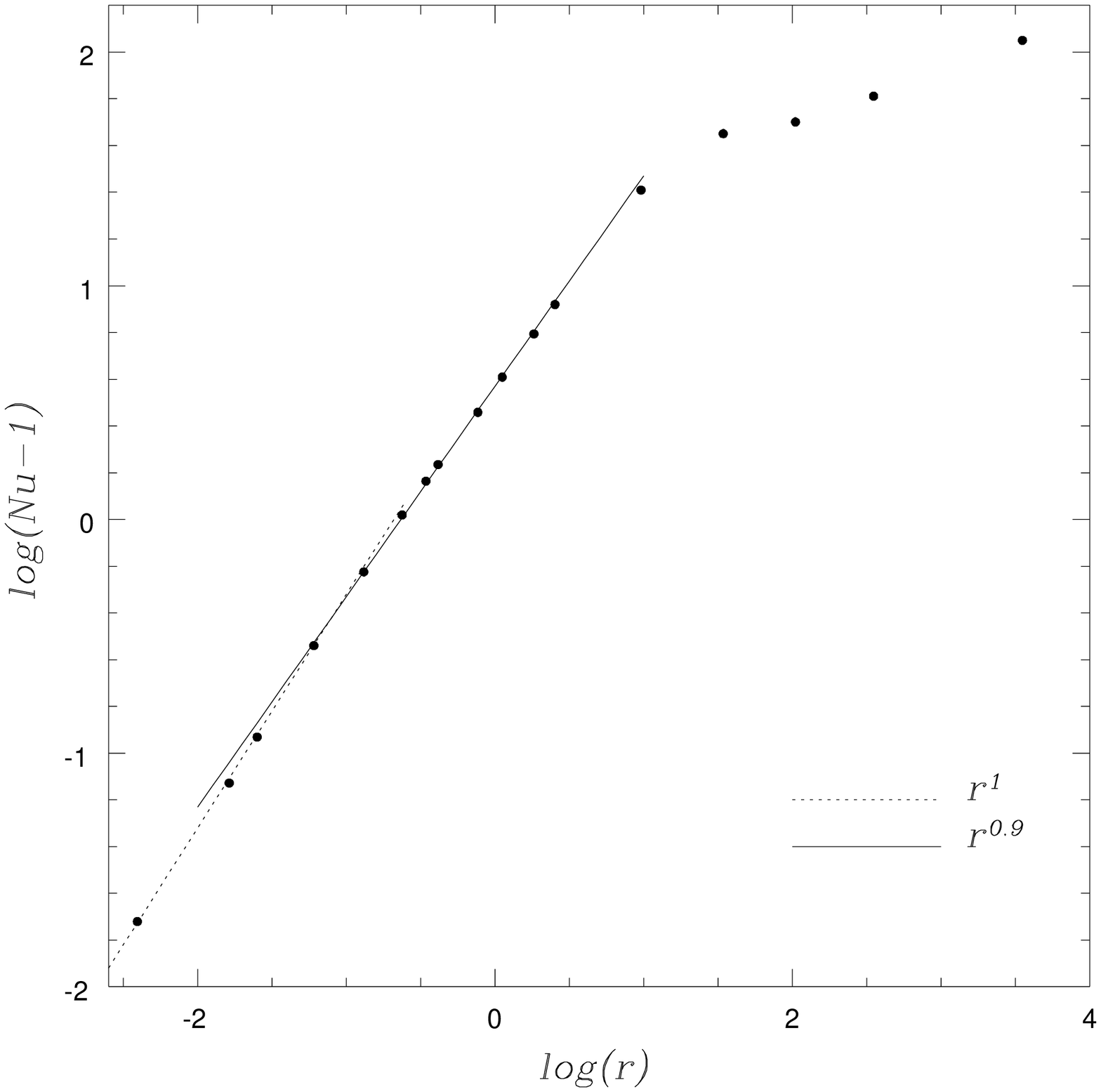,width=3.75in}}
\caption{}
\label{fig.plot_Nu_vs_Ra_I}
\end{figure}

\begin{figure}
\centerline{\psfig{file=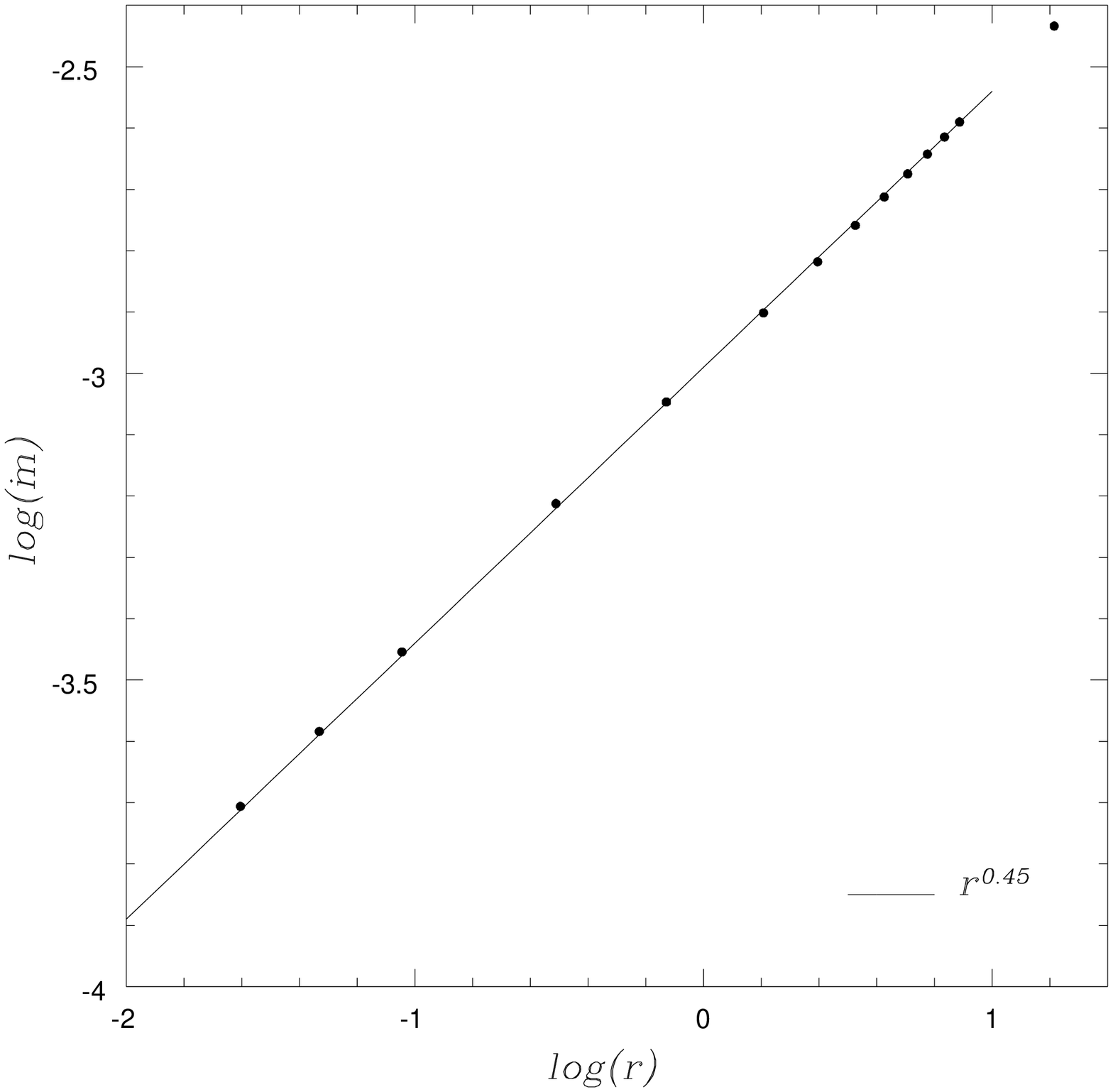,width=3.75in}}
\caption{}
\label{fig.plot_m_vs_Ra_II}
\end{figure}

\begin{figure}
\centerline{\psfig{file=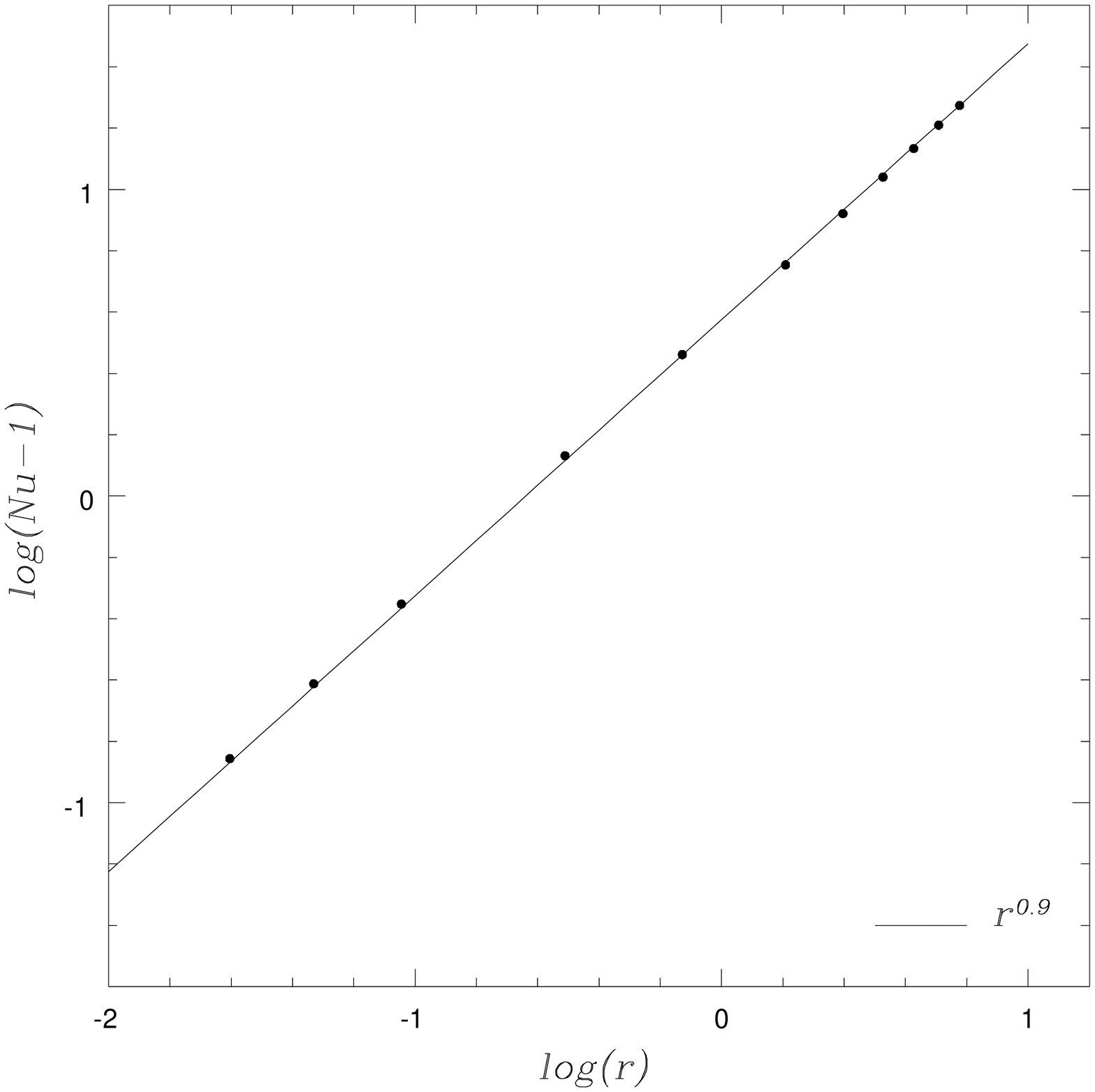,width=3.75in}}
\caption{}
\label{fig.plot_Nu_vs_Ra_II}
\end{figure}

\begin{figure}
\centerline{\psfig{file=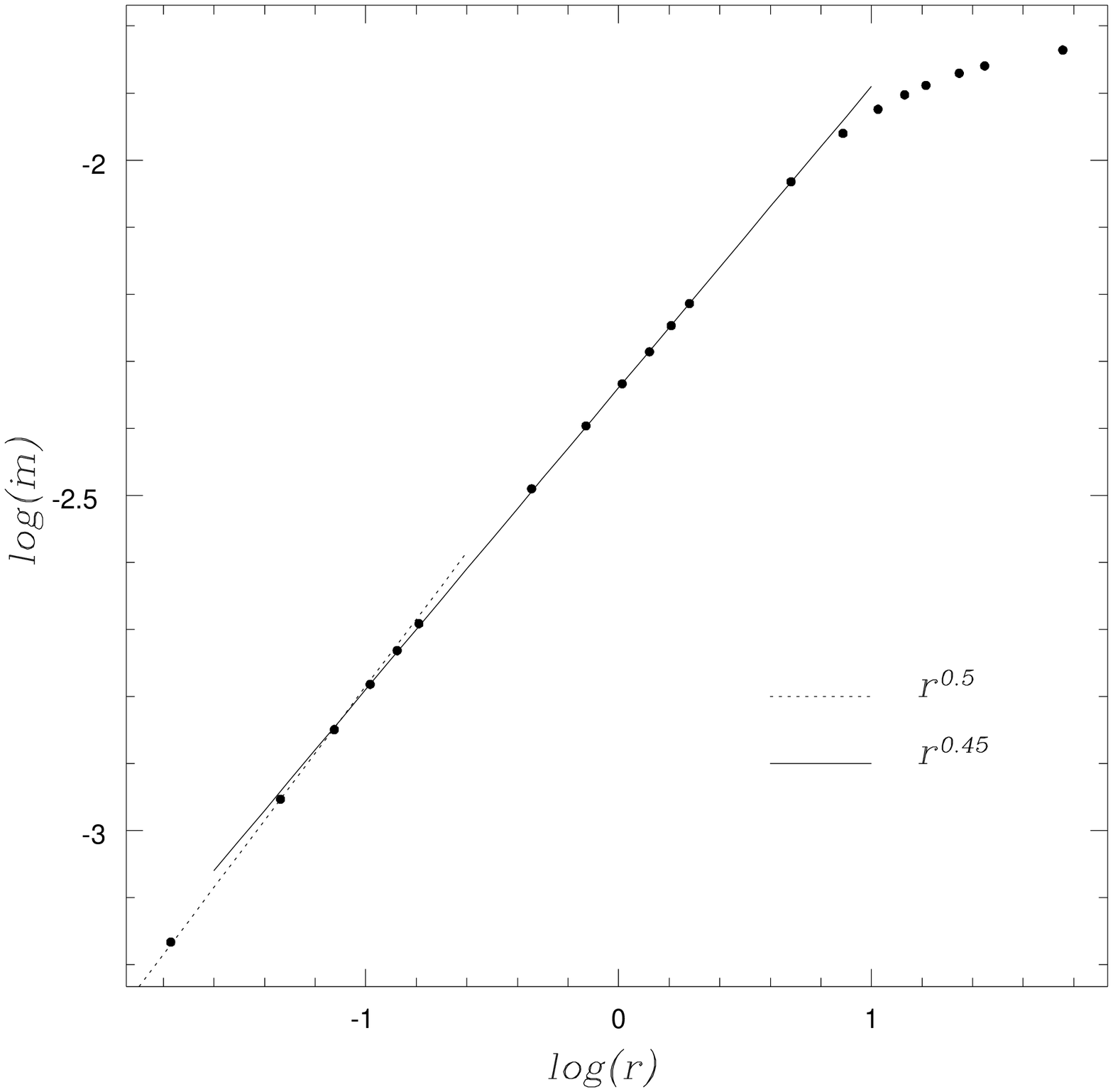,width=3.75in}}
\caption{}
\label{fig.plot_m_vs_Ra_III}
\end{figure}

\begin{figure}
\centerline{\psfig{file=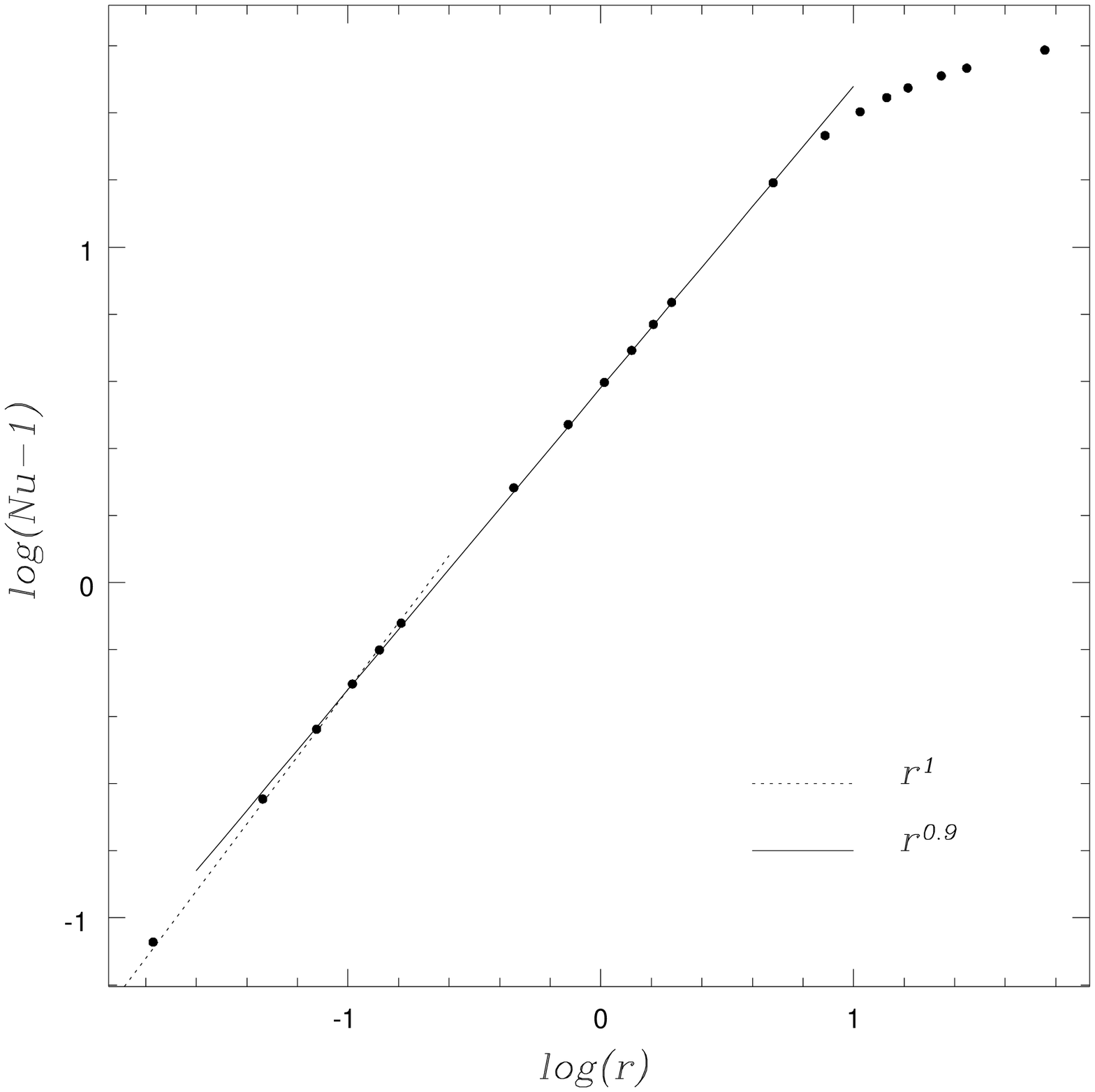,width=3.75in}}
\caption{}
\label{fig.plot_Nu_vs_Ra_III}
\end{figure}

\begin{figure}
\centerline{\psfig{file=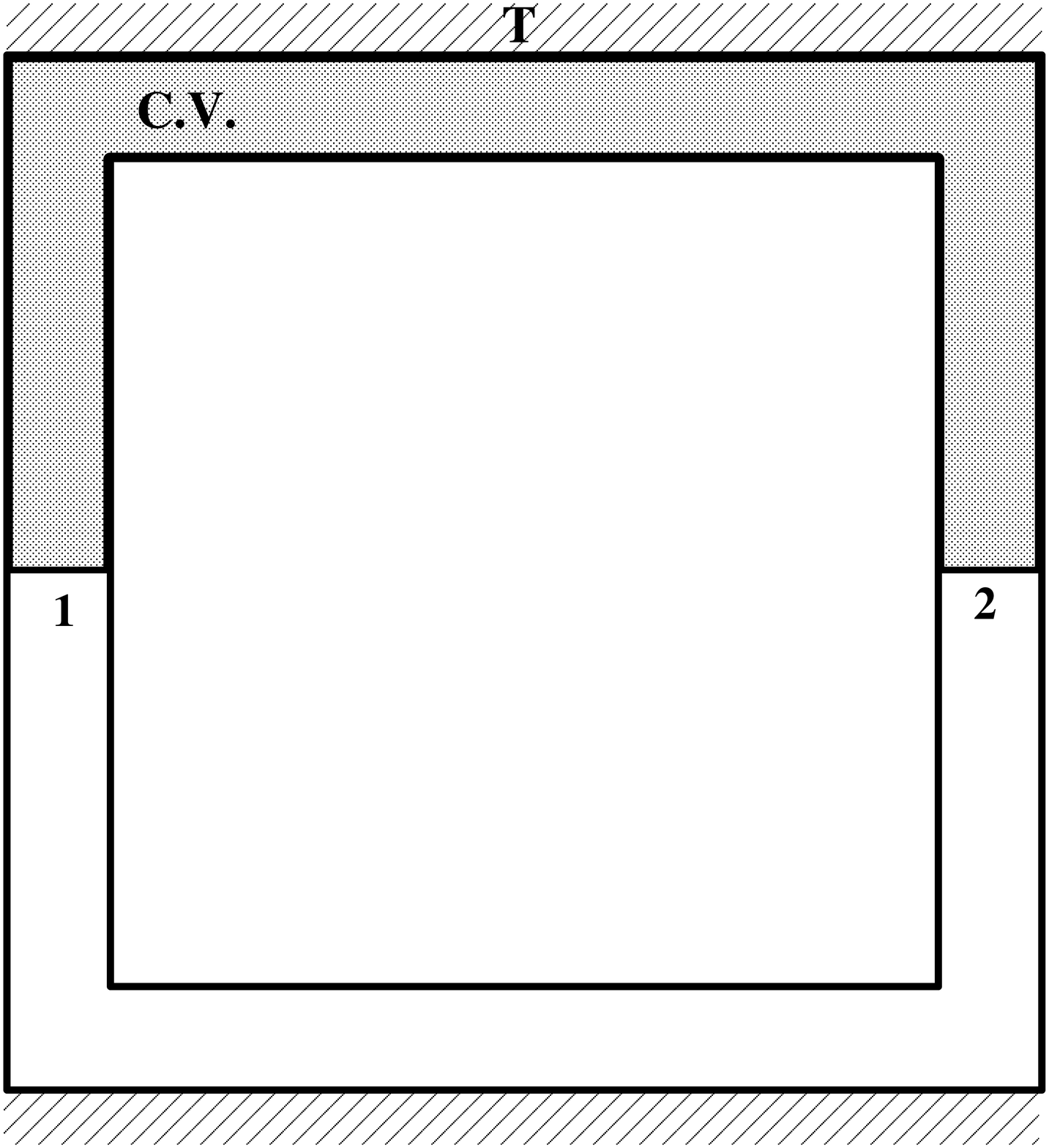,width=3.00in}}
\caption{}
\label{fig.CV}
\end{figure}

\begin{figure}
\centerline{\psfig{file=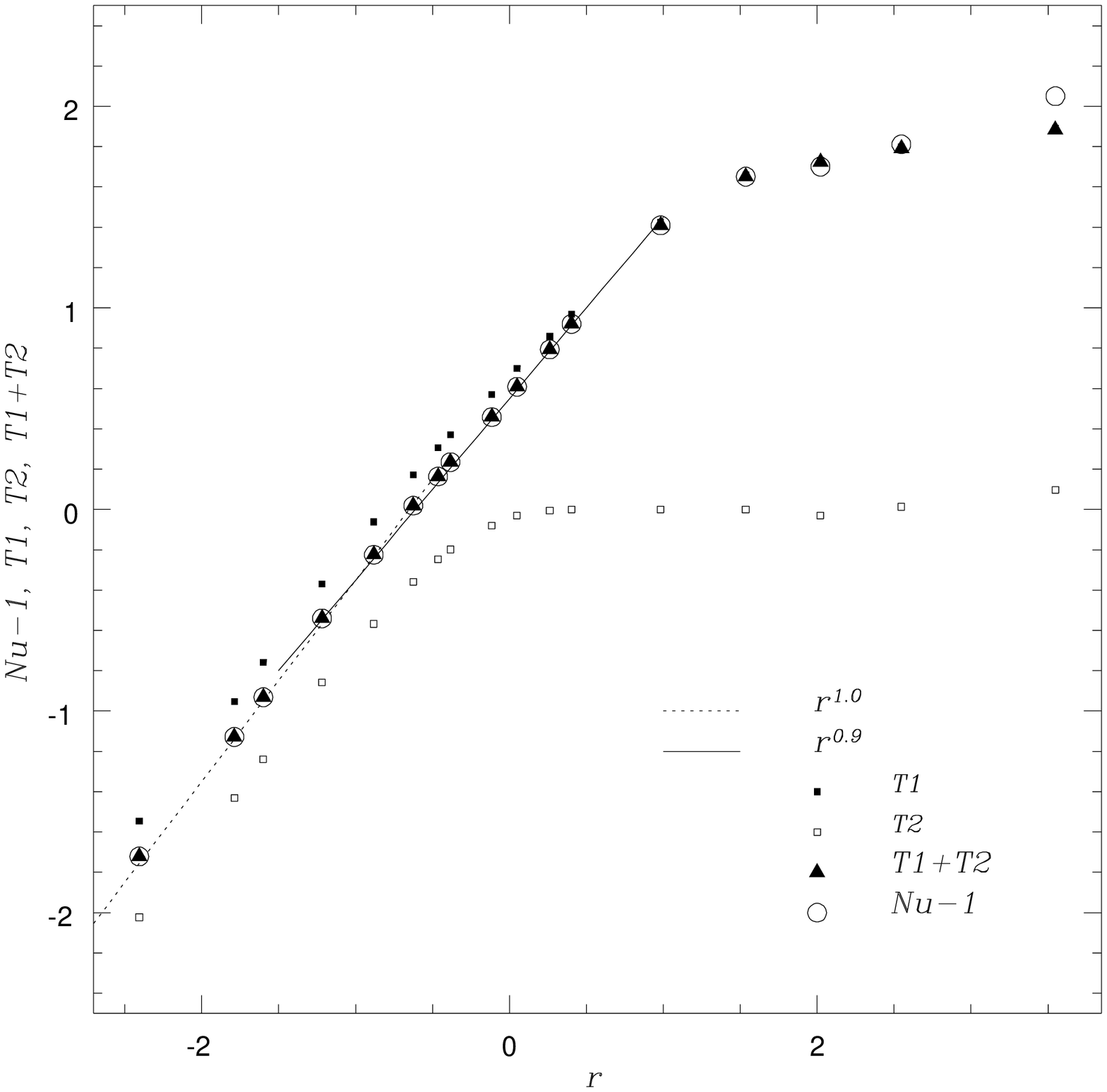,width=4.5in}}
\caption{}
\label{fig.lv_ltd_Nu}
\end{figure}

\begin{figure}
\centerline{\psfig{file=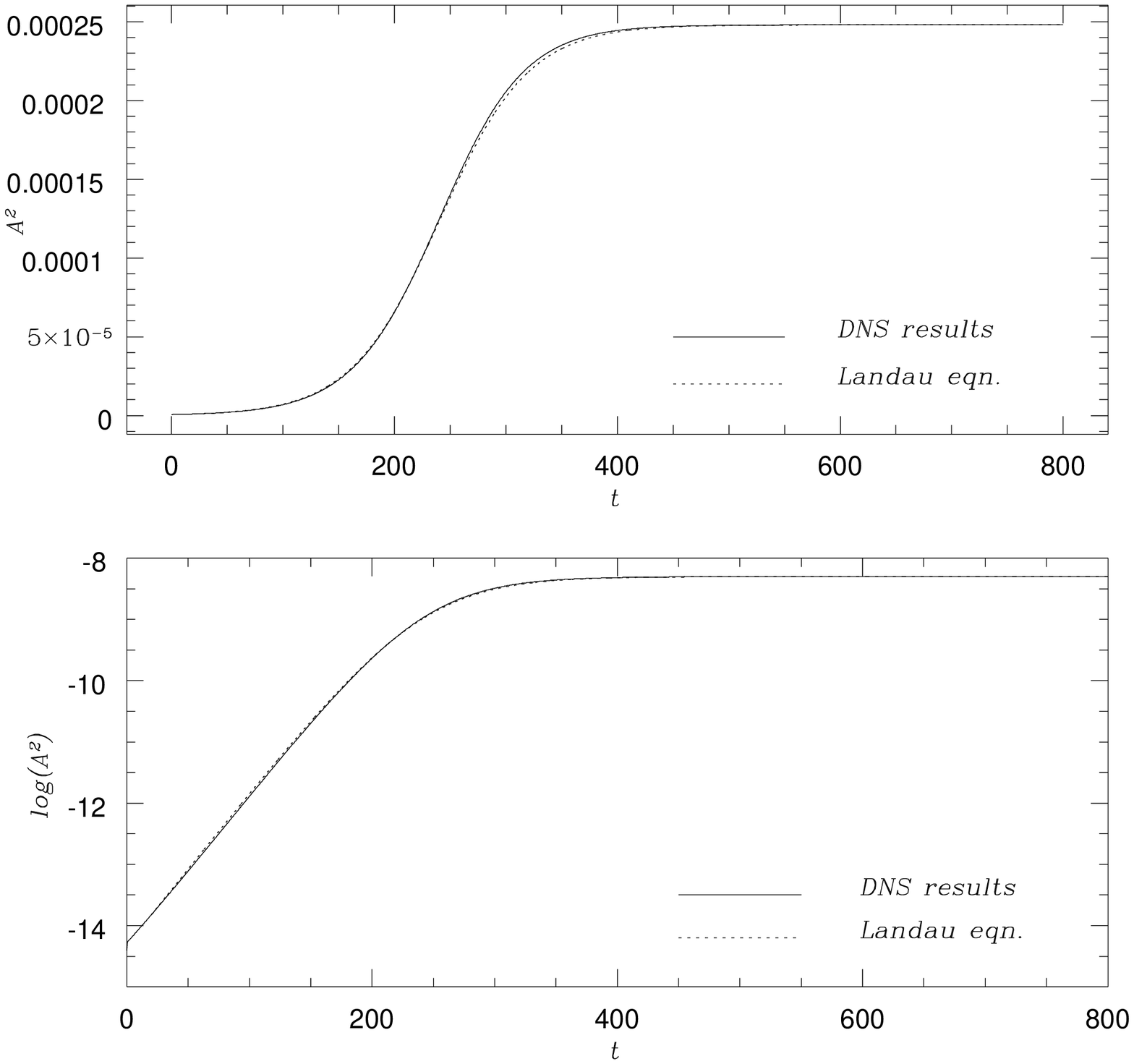,width=6.0in}}
\caption{}
\label{fig.plot_his_vs_Landau_Ra40000_III}
\end{figure}

\begin{figure}
\centerline{\psfig{file=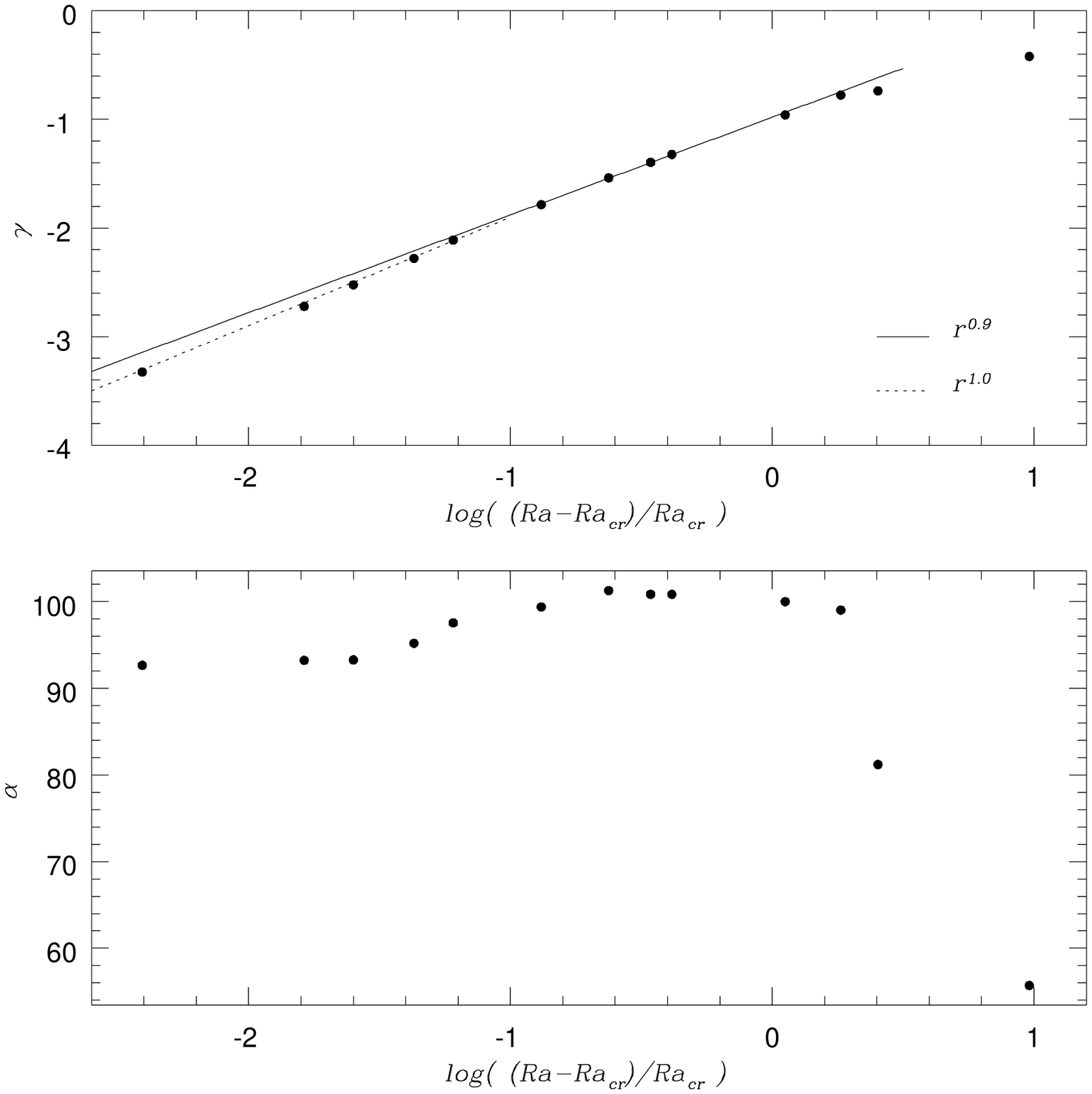,width=6.0in}}
\caption{}
\label{fig.plot_Landau_exponents_I}
\end{figure}

\begin{figure}
\centerline{\psfig{file=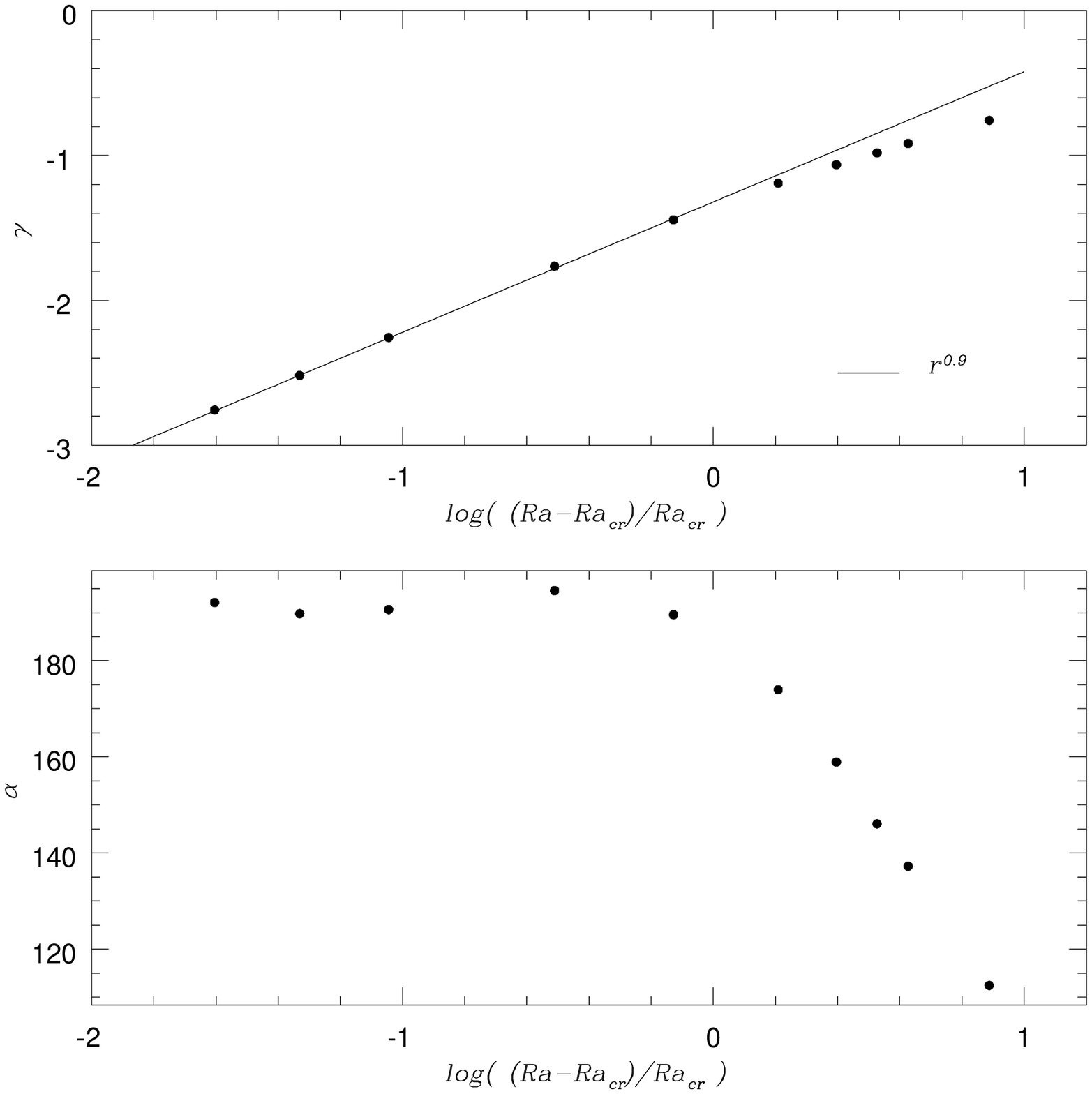,width=6.0in}}
\caption{}
\label{fig.plot_Landau_exponents_II}
\end{figure}

\begin{figure}
\centerline{\psfig{file=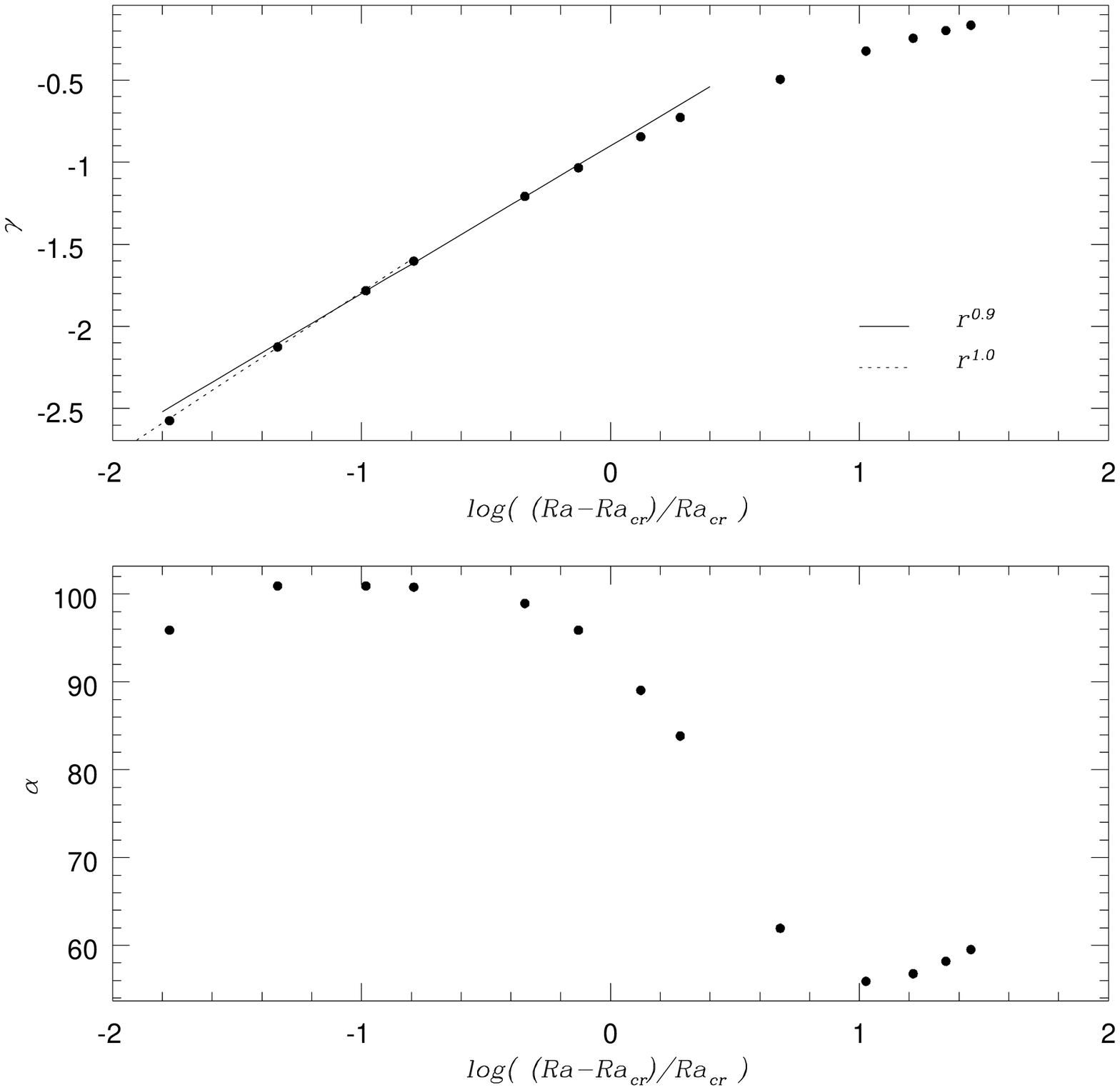,width=6.0in}}
\caption{}
\label{fig.plot_Landau_exponents_III}
\end{figure}

\begin{figure}
\centerline{\psfig{file=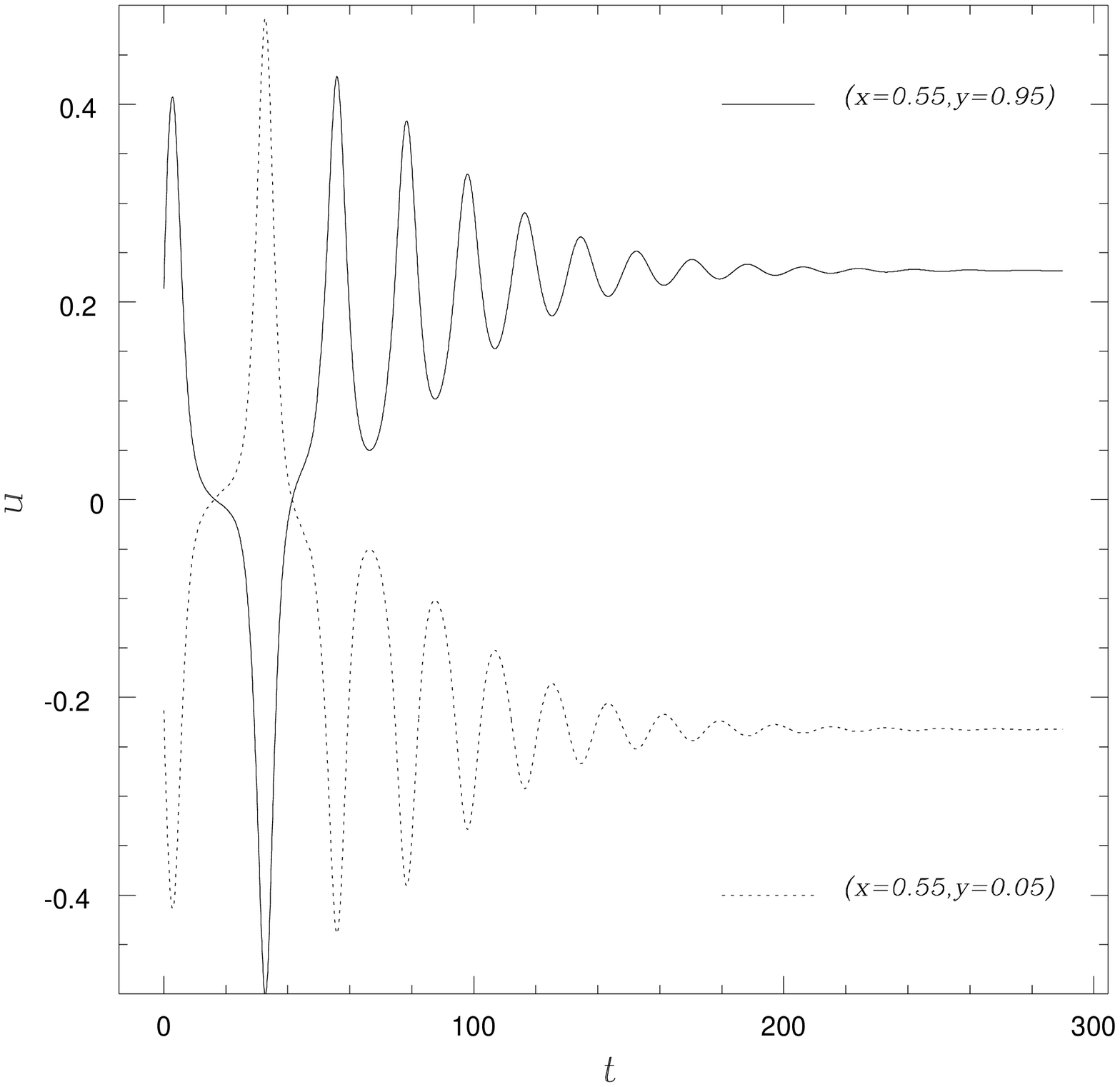,width=6.0in}}
\caption{}
\label{fig.Oscill}
\end{figure}

\begin{figure}
\centerline{\psfig{file=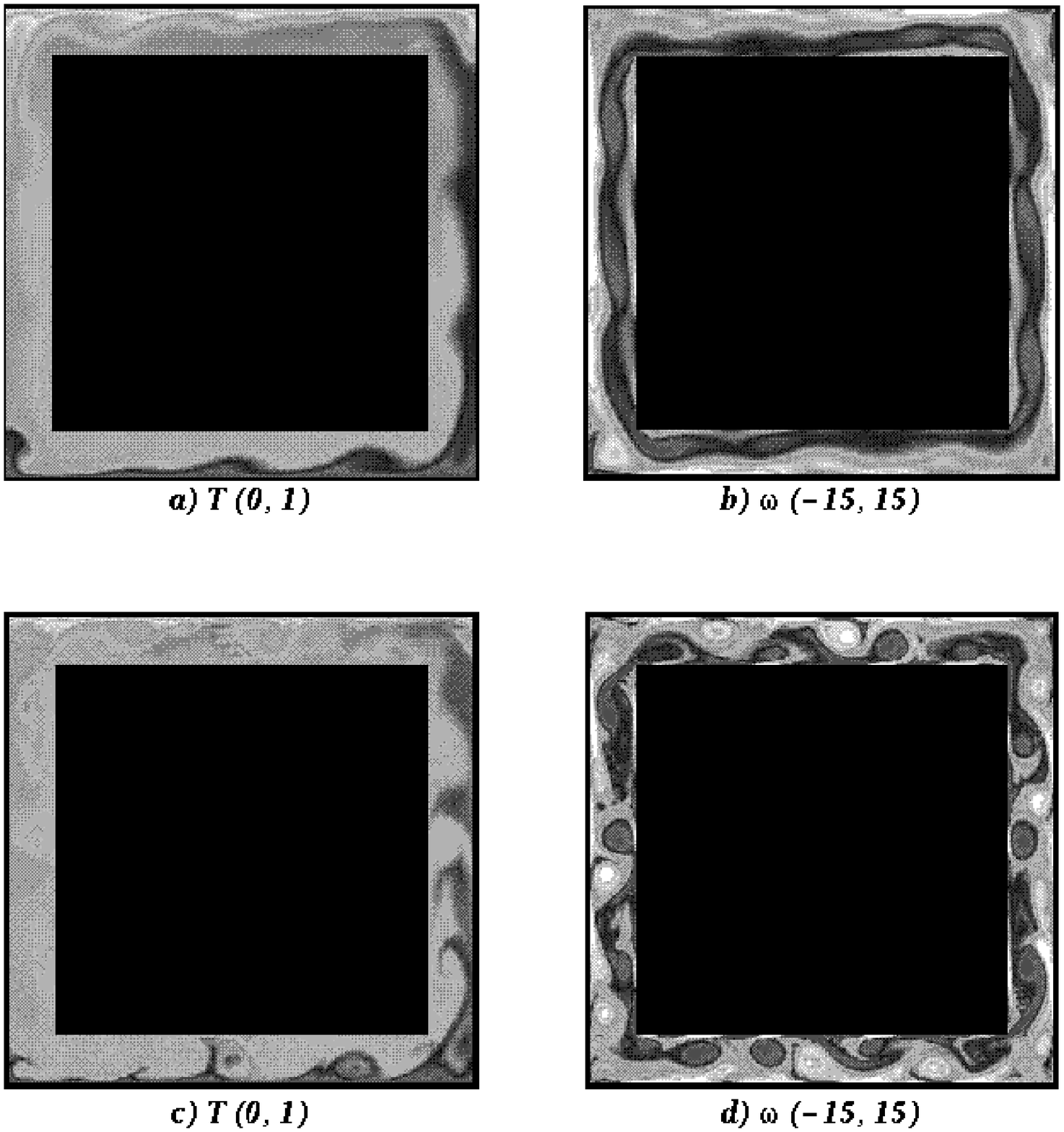,width=6.0in}}
\caption{}
\label{fig.isocontours_I}
\end{figure}

\begin{itemize}

\item Figure ~\ref{fig.config_all}: Geometric configuration of the three convection 
cells investigated

\item Figure ~\ref{fig.First_transition_Vort}: Formation of convection rolls for the 
three configurations I) Ra=10,000, II) Ra=100,000, III) Ra=30,000. Because this figure 
was generated by transformation of color plots to grey scale, the darkest tone does 
not correspond to the highest value.

\item Figure ~\ref{fig.A_vs_t_Ra30000_transition}: Total kinetic energy of the flow 
in time for $Ra=30,000$, case I, a) lin-lin plot, b) log-lin plot, which shows the linear 
theory regime

\item Figure ~\ref{fig.Second_transition_all}: a)Isocontours of the velocity in the 
$x-$direction $u$, b) in the $y-$direction $v$, c) temperature $T$ and d) vorticity 
$\omega$ for cases I, II, and III, and for $Ra=100,000$, $200,000$ and $200,000$, 
respectively. The minimum and maximum values of all variables are noted in each plot. 
Because this figure was generated by transformation of color plots to grey scale, the 
darkest tone does not correspond to the highest value.

\item Figure ~\ref{fig.profT_x0.05_vsRa}: Comparison of temperature profiles along $y$ 
for $x=0.05$, for case I with equation ~\ref{eqn.expT}.

\item Figure ~\ref{fig.plot_m_vs_Ra_I}: Logarithmic plot of the large scale average mass 
flux as function of $r=(Ra-Ra_{cr})/Ra_{cr}$, for case I
\item Figure ~\ref{fig.plot_Nu_vs_Ra_I}: Logarithmic plot of the $Nu$ number in the large 
scale flow regime ($Ra>Ra_{cr})$, as function of $r$, for case I

\item Figure ~\ref{fig.plot_m_vs_Ra_II}: Logarithmic plot of the large scale average mass 
flux as function of $r=(Ra-Ra_{cr})/Ra_{cr}$, for case II
\item Figure ~\ref{fig.plot_Nu_vs_Ra_II}: Logarithmic plot of the $Nu$ number in the large 
scale flow regime ($Ra>Ra_{cr})$, as function of $r$, for case II

\item Figure ~\ref{fig.plot_m_vs_Ra_III}: Logarithmic plot of the large scale average mass 
flux as function of $r=(Ra-Ra_{cr})/Ra_{cr}$, for case III
\item Figure ~\ref{fig.plot_Nu_vs_Ra_III}: Logarithmic plot of the $Nu$ number in the large 
scale flow regime ($Ra>Ra_{cr})$, as function of $r$, for case III

\item Figure ~\ref{fig.CV}: Control volume over which integration is performed

\item Figure ~\ref{fig.lv_ltd_Nu}: Variation of terms T1, T2, $Nu-1$, and T1+T2 (from equation 
~\ref{eqn.CV4}) with $r$.  Solid line is line with slope $0.9$ and dotted with $1.0$

\item Figure ~\ref{fig.plot_his_vs_Landau_Ra40000_III}: Time history of kinetic energy $A^2$, 
for case III and $r=0.16$ ($Ra=40,000$) Solid line is from DNS and dotted line from Landau 
amplitude equation

\item Figure ~\ref{fig.plot_Landau_exponents_I}: Logarithmic plot of the Landau constants 
$\gamma$ and $\alpha$ as functions of $r=(Ra-Ra_{cr})/Ra_{cr}$ in the large scale flow regime 
($Ra>Ra_{cr})$, for case I

\item Figure ~\ref{fig.plot_Landau_exponents_II}: Logarithmic plot of the Landau constants 
$\gamma$ and $\alpha$ as functions of $r=(Ra-Ra_{cr})/Ra_{cr}$ in the large scale flow regime 
($Ra>Ra_{cr})$, for case II

\item Figure ~\ref{fig.plot_Landau_exponents_III}: Logarithmic plot of the Landau constants 
$\gamma$ and $\alpha$ as functions of $r=(Ra-Ra_{cr})/Ra_{cr}$ in the large scale flow regime 
($Ra>Ra_{cr})$, for case III

\item Figure ~\ref{fig.Oscill}: Time history of $u-$velocity at the points shown, for case 
I and $Ra=10^7$

\item Figure ~\ref{fig.isocontours_I}: Isocontours of temperature $T$ (a,c) and vorticity 
$\omega$ (b,d) for case I and $Ra=10^8$ and $Ra=10^9$, respectively. The minimum and maximum 
values of all variables are noted in each plot. Because this figure was generated by transformation 
of color plots to grey scale, the darkest tone does not correspond to the highest value.

\end{itemize}

\newpage

\clearpage

\clearpage

\clearpage

\clearpage

\clearpage

\clearpage

\clearpage

\clearpage

\clearpage

\clearpage

\clearpage

\clearpage

\end{document}